\documentclass[aps,prl,twocolumn,showpacs,superscriptaddress,groupedaddress]{revtex4}
\usepackage{graphicx}  % needed for figures
\usepackage{dcolumn}   % needed for some tables
\usepackage{bm}        % for math
\usepackage{amssymb}   % for math
\usepackage{times}

\begin{document}

\title{Linear magnetoresistance caused by mobility fluctuations in the $n$-doped Cd$_3$As$_2$}

\author{A. Narayanan}
\affiliation{Clarendon Laboratory, Department of Physics, University of Oxford, Parks Road, Oxford OX1 3PU, U.K.}

\author{M. D. Watson}
\affiliation{Clarendon Laboratory, Department of Physics, University of Oxford, Parks Road, Oxford OX1 3PU, U.K.}

\author{S. F. Blake}
\affiliation{Clarendon Laboratory, Department of Physics, University of Oxford, Parks Road, Oxford OX1 3PU, U.K.}

\author{Y. L. Chen}
\affiliation{Clarendon Laboratory, Department of Physics, University of Oxford, Parks Road, Oxford OX1 3PU, U.K.}

\author{D. Prabhakaran}
\affiliation{Clarendon Laboratory, Department of Physics, University of Oxford, Parks Road, Oxford OX1 3PU, U.K.}

\author{B. Yan}
\affiliation{Max-Planck-Institut fur Chemische Physik fester Stoffe,  01187 Dresden,  Germany}

\author{N. Bruyant}
\affiliation{Laboratoire National des Champs Magn´etiques Intenses (CNRS), 31077 Toulouse, France}

\author{L. Drigo}
\affiliation{Laboratoire National des Champs Magn´etiques Intenses (CNRS), 31077 Toulouse, France}

\author{I. I. Mazin}
\affiliation{$^{2}$Code 6393, Naval Research Laboratory,
Washington, D.C. 20375, USA}

\author{C. Felser}
\affiliation{Max-Planck-Institut fur Chemische Physik fester Stoffe,  01187 Dresden,  Germany}

\author{T. Kong}
\affiliation{Ames Laboratory, Iowa State University, Ames, Iowa 50011, USA}
\affiliation{Department of Physics and Astronomy, Iowa State University, Ames, Iowa 50011, USA}

\author{P. C. Canfield}
\affiliation{Ames Laboratory, Iowa State University, Ames, Iowa 50011, USA}
\affiliation{Department of Physics and Astronomy, Iowa State University, Ames, Iowa 50011, USA}

\author{A. I. Coldea}
\email[corresponding author:]{amalia.coldea@physics.ox.ac.uk}
\affiliation{Clarendon Laboratory, Department of Physics, University of Oxford, Parks Road, Oxford OX1 3PU, U.K.}

\begin{abstract}
Cd$_{3}$As$_{2}$ is a candidate three-dimensional Dirac semi-metal which has exceedingly high mobility and non-saturating linear magnetoresistance that may be relevant for future practical applications. We report magnetotransport and tunnel diode oscillation measurements on Cd$_{3}$As$_{2}$, in magnetic fields up to 65~T and temperatures between 1.5~K to 300~K. We find the non-saturating linear magnetoresistance persist up to 65~T and it is likely caused by disorder effects as it scales with the high mobility, rather than directly linked to Fermi surface changes even when approaching the quantum limit. From the observed quantum oscillations, we determine the bulk three-dimensional Fermi surface having signatures of Dirac behaviour with non-trivial Berry's phase shift, very light effective quasiparticle masses and clear deviations from the band-structure predictions. In very high fields we also detect signatures of large Zeeman spin-splitting ($g \sim 16$).

\end{abstract}

\pacs{75.47.-m, 71.18.+y, 74.25.Jb}
\date{\today}
\maketitle

A three-dimensional (3D) Dirac semi-metal is a three-dimensional analogue of graphene,
 where the valence and conduction bands touch at discrete points in reciprocal space with a linear dispersion.
These special points are protected from gap formation by crystal symmetry
and such a topologically non-trivial band structure may harbour  unusual electronic states.
A Dirac semi-metal may be tuned to attain a Weyl semi-metal phase through breaking of inversion or time reversal symmetry \cite{theory}. Alternatively, if the symmetry protection from gapping is removed a three dimensional topological insulator could be stabilized on the surface \cite{theory}.
3D Dirac semi-metals are rare and an opportunity to realize such a state in Cd$_{3}$As$_{2}$ has
generated a lot of interest. Surface probes, such as ARPES and STM \cite{arpes1,arpes2,arpes3,stm},
found that the linear dispersion extends up to
high energy 200-500~meV, strongly dependent on the
cleavage directions \cite{arpes4}.
Furthermore, the large non-saturating linear
magnetoresistance (MR) found in Cd$_{3}$As$_{2}$ \cite{luli,onglmr}
in high mobilities samples was assigned to the lifting of protection against
backscattering caused by possible field-induced Fermi surface changes
\cite{luli,onglmr}.

In this paper we report a magnetotransport study in high magnetic fields up to 65~T of $n$-doped Cd$_{3}$As$_{2}$
beyond the quantum limit that reveal no discernable Fermi surface change except
those caused by the large Zeeman splitting.
  We observe Shubnikov-de Haas (SdH) quantum oscillations that allow us to characterize the three-dimensional
  Fermi surface and its relevant parameters. The observed linear MR
 in ultra-high magnetic fields and the values of the linear magnetoresistance are closely linked to the mobility field scale.
  This suggests that the unconventional, non-saturating, large and linear magneto-resistance
 in our electron-doped crystals of Cd$_{3}$As$_{2}$
 is likely to originate from mobility fluctuations caused by As vacancies.
 %described by a model of Parish and Littlewood \cite{parish}, rather than the Abrikosov model \cite{abrikosov}
% of quantum magnetoresistance or induced by Fermi surface
%changes in magnetic field \cite{luli,onglmr}.
We also discuss the deviations of experiments from the standard density functional theory
(DFT) calculations. %that predict (in the absence of electronic correlations)
%that a Lifshits transition occurs a much lower energies than experimentally found ($\sim $ 5 meV), leading to developing
%of parabolic bands dispersion in some directions.

%Furthermore the MR of a Dirac system is believed to contain various exotic effects that are currently swamped by the linear MR \cite{theory}. Thus, it becomes crucial to determine the source of the linear MR, whether it is of Dirac origin or not, and to find means to address the bulk electron dispersion directly.

{\it Methods}
Crystals of Cd$_{3}$As$_{2}$ were grown both by solid state reaction
and solution growth from Cd-rich melt due
to it very narrow growth window \cite{cava,Canfield1992}. X-ray diffraction
show that our single crystals of Cd$_{3}$As$_{2}$ crystallize
in the tetragonal symmetry group $I4_1/acd$ with lattice parameters $a$=12.6595(6)~\AA~ and $c$=25.4557(10)~\AA,
  cleaving preferentially in the (112) plane,
in agreement with previous studies \cite{cava} [see Supplementary Material (SM)]. Band structure calculations
were performed with Wien2K including the spin-orbit coupling \cite{Blaha2001} using
the structural details from Ref.\cite{cava}.
 We have performed magnetotransport measurements in the standard Hall and resistivity configuration using a low frequency lock-in technique by changing the direction of the magnetic field, ${\bf B}$,
to extract the symmetric ($\rho_{xx}$) and the anti-symmetric ($\rho_{xy}$) component of the resistivity tensor, respectively.
The transverse magnetoresistance ($I \perp B $) was measured for different orientations,  $\theta$
being the angle between ${\bf B}$ and the normal to the (122) plane.
Measurements were conducted on three different batches ($a$, $b$ and $c$), mostly
on crystals from batch $a$ ($S^a_{1}$, $S^a_{2}$, etc.)
having the lowest carrier concentration.
Measurements were performed at low temperatures
(1.5~K) in steady fields up to 18~T in
Oxford and in pulsed fields up to
65~T at the LNCMI,
%(Laboratoire National des Champs Magnetiques Pulses,)
Toulouse.
 We also measured skin depth in pulsed fields
using a tunnel diode oscillator technique (TDO) by
recording the change in frequency of an LC tank circuit with
the sample wound in a copper coil, reported data
being corrected for the magnetoresponse of the empty coil.

\begin{figure}[ht!]
\centering
  \includegraphics[width=8cm]{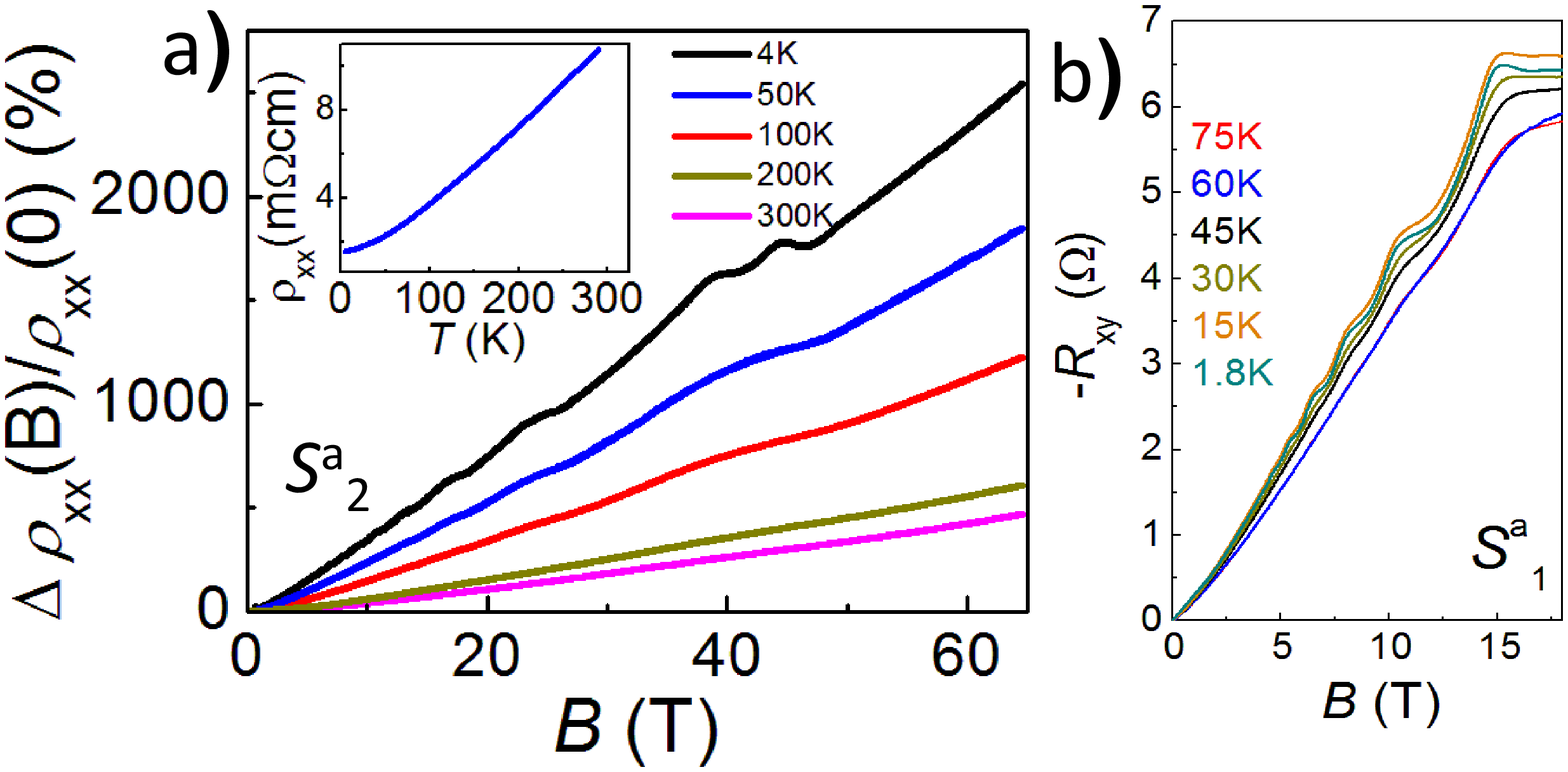}
  \includegraphics[width=8cm]{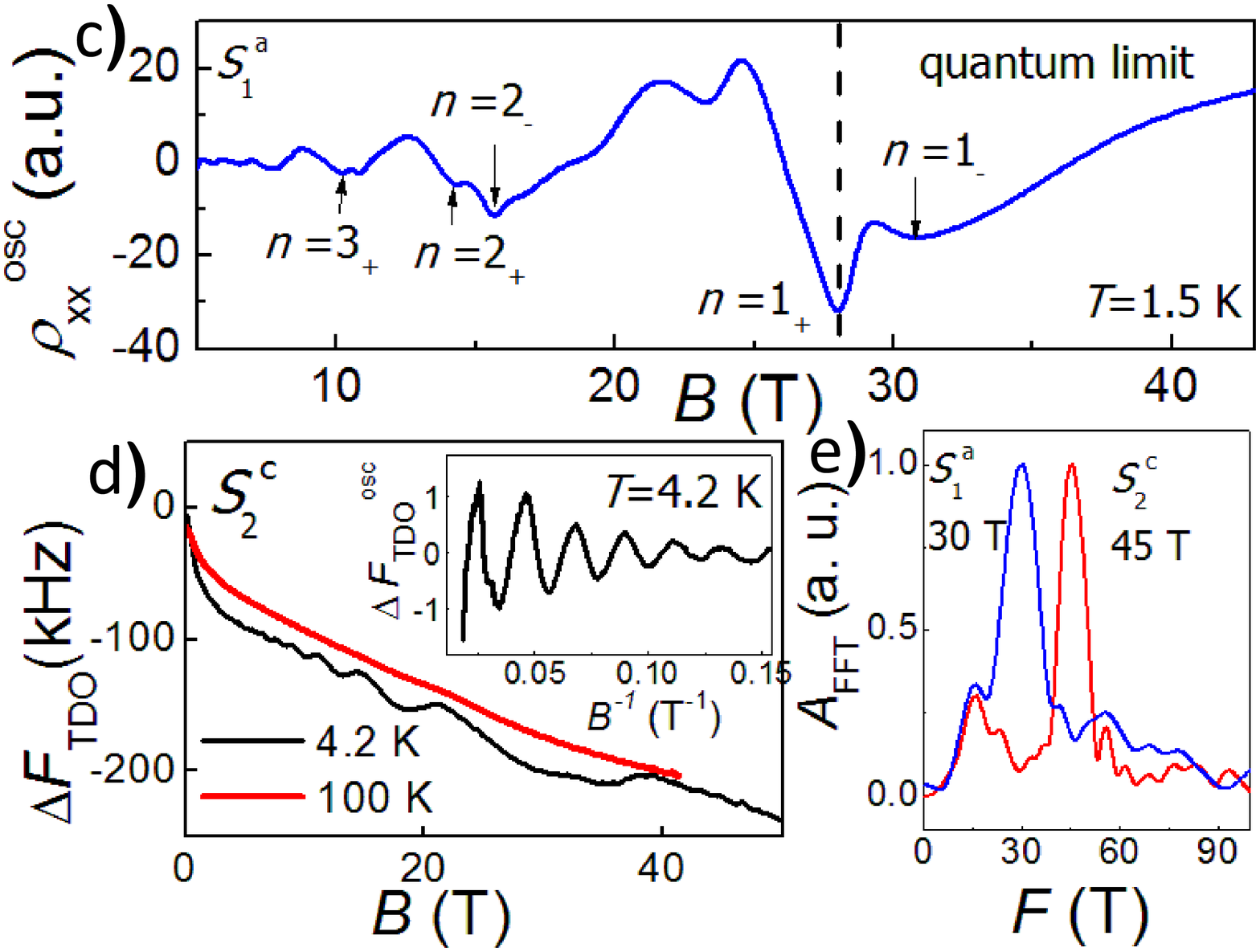}
  \caption{High magnetic field data. a) Field dependence of $\rho_{xx}$
  and the relative change in magnetoresistance, $\Delta \rho_{xx}$/$\rho_{xx}$(0) for sample $S^a_{2}$ up to 65~T for temperatures between 4~K and 300~K. b) Field dependence of Hall resistance, $R_{xy}$, for sample $S^a_{1}$ up to 18~T. c) The oscillatory part of symmetrized $\rho_{xx}$ for $S^a_{2}$ reaching the quantum limit. The arrows indicate the position of different spin-split Landau level crossing the Fermi level. d) The field dependence of resonant frequency, $\Delta F_{\rm TDO}$, of a tunnel diode oscillator for sample $S^c_{2}$ up to 55~T. The inset shows the oscillatory part of $\Delta F_{\rm TDO}$. e) FFT frequencies corresponding to oscillatory signal in c) and d).}
  \label{fig1}
  \end{figure}

Fig.~\ref{fig1}a) shows the magnetoresistance, $\Delta \rho_{xx}$ (B)/$\rho_{xx}$(0)
as a function of magnetic field up to 65~T for sample $S^a_{2}$
at fixed temperatures between
 4~K and 300~K. The MR is linear and unusually large, $\sim 20000\%$ 
and shows a strong temperature dependence.
Both the resistance and the magnetoresistance
change by a factor of 5 from 300~K to 4~K  [inset of Fig.~\ref{fig1}a]) and
the link between these two quantities
 will be discussed in detail later.
Fig.~\ref{fig1}b) show the Hall component, $\rho_{xy}$
up to 18~T for $S^a_{1}$ up to 75~K (raw data also in SM).
 Quantum oscillations are discernible, on a highly linear background, from as low
 as 3~T with a characteristic frequency varying
 for different samples between 30 to 50~T, shown in Fig.\ref{fig1}c-e) and listed in Table~1.
  Spin-splitting effects
 are evident in very high magnetic fields approaching the quantum limit ($n$=1) in Fig.\ref{fig1}c.
%\footnote{By plotting the derivative $\frac{\mathrm{d} log(R(B))}{\mathrm{d} log(B)}$, the exponent $\alpha$ in R $\propto B^{\alpha}$ can be checked to be 1 $\pm 0.1$.}.
%
The field dependence of resonant frequency
from TDO measurements for sample $S^c_2$ is shown
in Fig.~\ref{fig1}d)
together with subtracted quantum oscillations.
This frequency variation, $\Delta F_{TDO}$,
tracks the change in impedance of the coil
and is a measure of the skin depth of the sample,~$\delta \propto \rho_{xx}^{0.5}$.

\begin{figure}[htb]
\centering
  \includegraphics[width=8cm,clip=true]{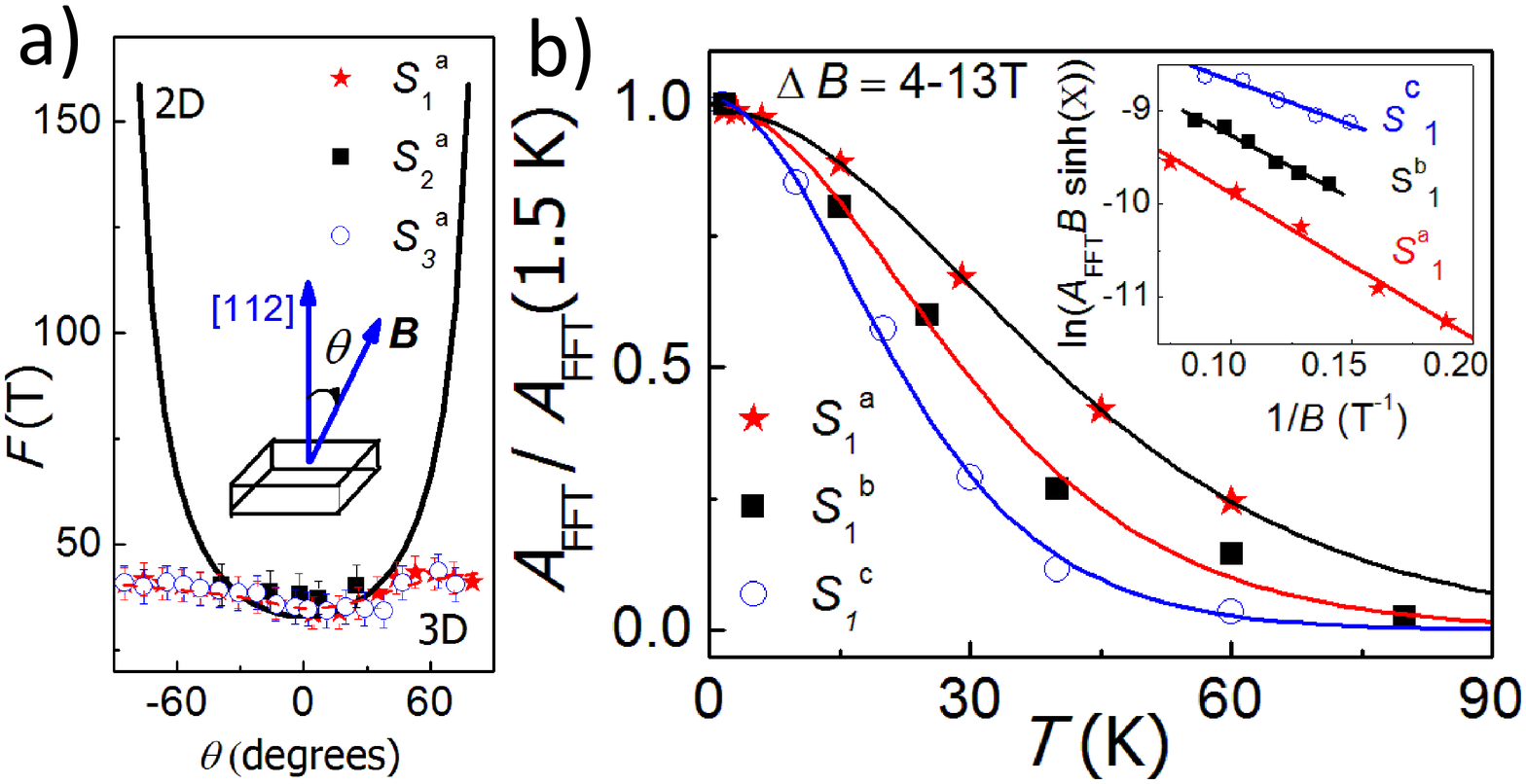}
  \includegraphics[width=8cm,clip=true]{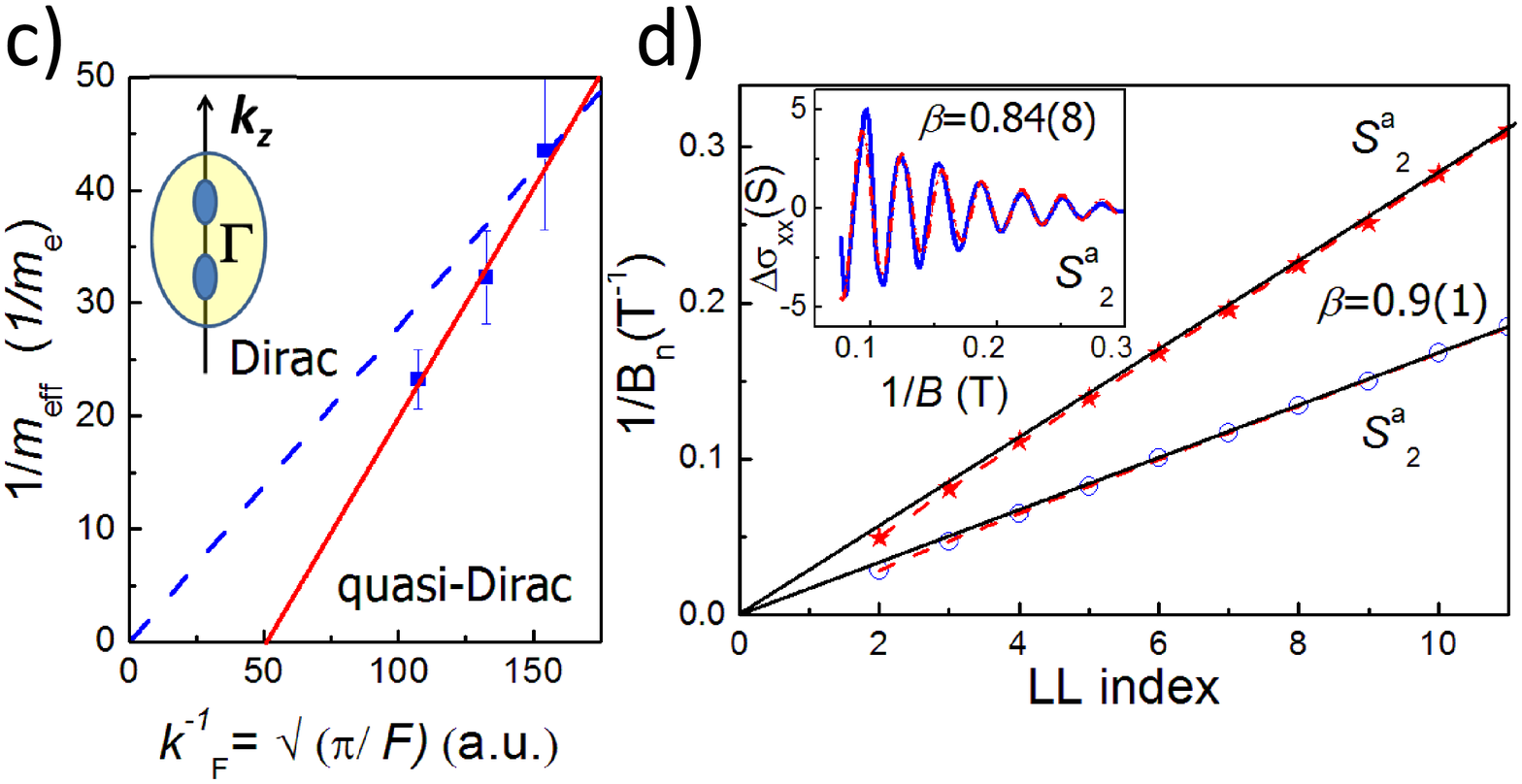}
 \caption{Fermi surface parameters.  a)  The angular dependence
  of SdH oscillation frequencies away from the (112) plane. The solid
  line is the expectation for a two-dimensional Fermi surface.
  b) The temperature dependence of the oscillation amplitude that gives $m_{eff}$ for
  different samples ($S^{a}_1$, $S^{a}_2$ and $S^{a}_3$). Inset shows
   the Dingle plots of the FFT amplitude for samples from different batches (a, b and c).
  c) Extracting the Fermi velocity from a linear fit of
   $1/m_{eff}$ versus $\sqrt{(\pi/F)}$ (in atomic units, a.u.),
   as described in main text (solid line). The dashed line indicates
   the expected behaviour for a perfect Dirac system. The inset show a schematic
   Fermi surface before and after the Lifshitz transition.
     d) Index plot to extract
    the Berry's phase $\beta = 0.9(1)\pi$ (as detailed in the text).
    Inset show quantum oscillations in conductivity $\sigma_{xx}$  fitted to the
    Lifshitz-Kosevich formula (dashed line) \cite{Shoenberg1984} with a phase of $\beta$ = 0.84(8).}
  \label{fig2}
  \end{figure}

{\it Quantum oscillations}
The quantum oscillations in conductivity are
given by $\Delta \sigma _{xx}\propto  \cos (2\pi \left [ \frac{F}{B} -\frac{1}{2} + \beta   \right ])$,  where
$\beta$ is the Berry's phase and $F$ is the SdH frequency of the oscillations, corresponding
  to an extremal area of the Fermi surface perpendicular to the magnetic field, $B$.
Fig.~\ref{fig2}a) shows the angular dependence of SdH frequencies
by rotating away from the (112) plane for different samples.
The SdH frequencies show very little variation as a function
of the orientation in magnetic field,
 from 31~T to 45~T for sample $S^a_1$ (see also Table~1).
 This behaviour is expected for a three-dimensional elliptical Fermi surface
 with $k_F$ vector, extracted
 from Osanger relationship $F=\hbar \pi k^2_F/(2 \pi e)$, and
 varying between $k_F = 0.03-0.04 $ \AA$^{-1}$.
  %The dotted line is a fit to a cosine dependence expected for 2-D surface states showing clear %divergence from the data at large angles.
  This values give a very small carrier concentration of $n_{\rm SdH}=1.0(2) \times 10^{18} $cm$^{-3}$,
  consistent with that from Hall measurements $n_{\rm Hall}$=$1.8 \times 10^{18} $cm$^{-3}$
  [extracted from $R_{xy}$ in Fig.\ref{fig1}b],
  assuming two elliptical pockets as shown in Table~1.
A Lifshitz transition as a function of doping
occurs from two small elliptical Fermi surfaces centered at the Dirac node
($k_z \sim 0.15$\AA$^{-1}$ away from $\Gamma$) \cite{arpes3} to
 a larger merged elliptical Fermi surface centered now at $\Gamma$ (see SM).
 Band structure calculations suggest that
 this transition should occur very close to the Fermi level ($\sim 40$~meV),
 whereas in the surface experiments is not seen up to
300~meV  \cite{stm,arpes3} (see inset Fig.2c and SM).
This is a rather surprising discrepancy between the band structure and experiments
and it will required further understanding.

The temperature dependence of the amplitude of the quantum oscillations
up to 90~K  can be used to extract the values of the effective
cyclotron mass $m_{eff}$,
using the standard  Lifshitz-Kosevich formalism \cite{Shoenberg1984},
 $T / \sinh ( 2 \pi^2 T m_{eff} / \hbar eB) $, which
also holds for the Dirac spectrum \cite{Sharapov2004},
as shown in Fig. \ref{fig2}b).
 For parabolic bands, one would expect $m_{eff}$
 to be constant as a function of doping, while for Dirac bands
$m_{eff} = \hbar k_F / v_{F}$.
The measured effective mass extracted for our samples from different
batches vary from 0.023 to 0.043~$m_e$, increasing with $F$
and the corresponding carrier concentration, $n_{SdH}$,
 as listed in Table~1. This suggests a deviation from a parabolic band dispersion
whereas the high mobility values found in Cd$_3$As$_2$ points usually towards a linear dispersion.
Having samples with different concentrations, one could attempt to extract the Fermi velocity, $v_{F}$,
directly from the slope of $ 1/m_{eff}$ versus  $k_F^{-1}$, shown in Fig. 2c,
which gives a finite intercept suggesting a departure from a perfect Dirac behaviour
[possibly linked to band structure effects that show hole-like bending towards $\Gamma$ (see SM)].
The estimation of $v_{F}\approx 4\times10^{6} $ is similar to those extracted from ARPES, $0.8-1.5\times 10^{6}$ m/s \cite{arpes1,arpes3}, with deviations caused by orbitally-averaged effects (see also Table~1).
  We have also extracted the values of $g$-factor from the spin-split oscillations
  visible at  high fields [see Fig.~\ref{fig1}c)],
   corresponding to the spin-up and spin-down Landau levels
($\pm g \mu_B B$) that cross the Fermi level
 and give a large value of $g \sim 16(4)$, consistent with previous reports \cite{oldgfactor1,oldgfactor2}.

 The Berry's phase, $\beta$,  can take values of $\beta = 0$ for parabolic dispersion  and $\beta = \pi$ for a Dirac point \cite{Taskin2011}.
   To extract the Berry's phase, we use the conductivity, $\sigma_{xx}$, by measuring both $\rho_{xx}$ and $\rho_{xy}$ simultaneously (see SM) and inverting the resistivity tensor, as shown in the inset of Fig.\ref{fig2}d). The direct fit
  of  $\Delta \sigma _{xx}$ gives a value of $\beta =0.84(8) \pi$, in agreement
with previous reports, as shown in Table~1. Another method to extract $\beta$
 is given by the linear intercept of an index plot of the conductivity minima versus inverse magnetic field;
 in the low-field region (from $n$=4) that gives $\beta =0.8(1)\pi $ [solid line in Fig.\ref{fig2}d]),
 whereas in high magnetic fields the position of the minima are affected by the spin-splitting  and a non-linear fan diagram analysis detailed in Ref.\cite{Taskin2011,mckenzie} gives $\beta =0.9(1)\pi $ [dashed line in Fig.\ref{fig2}d)].

{\it Scattering}
The field dependence of the amplitude of quantum oscillations at fixed temperatures
[inset Fig.\ref{fig2}b)] gives access to the Dingle temperature,
which is a measure of the field dependent damping of the quantum oscillations
due to impurity scattering.
For sample $S^a_{1}$ the  quantum scattering time given
by $\tau_q = \hbar$/($2\pi k_{B}T_{\rm D}$),
%is $\tau_q = 1 \times 10^{-13}s$,
corresponds to a quantum mobility of  $\mu_{q} \sim 6000$ cm$^2$/Vs
  and mean free path  $\ell_q=v_{F}\tau_q$ of $\sim 122(8) $ nm. These values are in good
 agreement to some of the
reports for single crystals and thin films, as shown in Table~1.
Another way to estimate the mobility is to apply a simple Drude model to the Hall and resistivity data. Using the carrier concentration estimated from the Hall effect $n_{H}= 1.8 \times 10^{18} $cm$^{-3}$
and $\rho_{xx0} = 42 \mu \Omega $cm
for sample $S^a_1$ (shown in SM),
 the classical mobility from $1/\rho_{xx} =n_{H} \mu_c e$ is $\mu_c$=80,000~cm$^{2}$/Vs, a factor up to 13 larger than the mobility from quantum oscillations, $\mu_q$. This difference in the two mobilities is common as they measure different scattering processes. The SdH estimated mobility is affected by  all processes that cause the Landau level broadening, {\it i.e} quantum scattering time, $\tau_q$, measures how long a carrier stays in a momentum eigenstate whereas the classical Drude mobility is only affected by scattering processes that deviate the current path, {\it i.e} the classical scattering time (transport time) is a measure of how long a particle moves along the applied electric field gradient.
Thus, the quantum mobility is susceptible to small angle and large angle scattering, while the transport (classical) mobility is susceptible only to large angle scattering. The ratio $\mu_c$/$\mu_q$ is a measure of the relative importance of small angle scattering; Table~1 suggests that small angle scattering dominates in all our samples,
in particular for lower doping $n_{\rm SdH}$.

\begin{figure}[htb]
\centering
  \includegraphics[width=8cm]{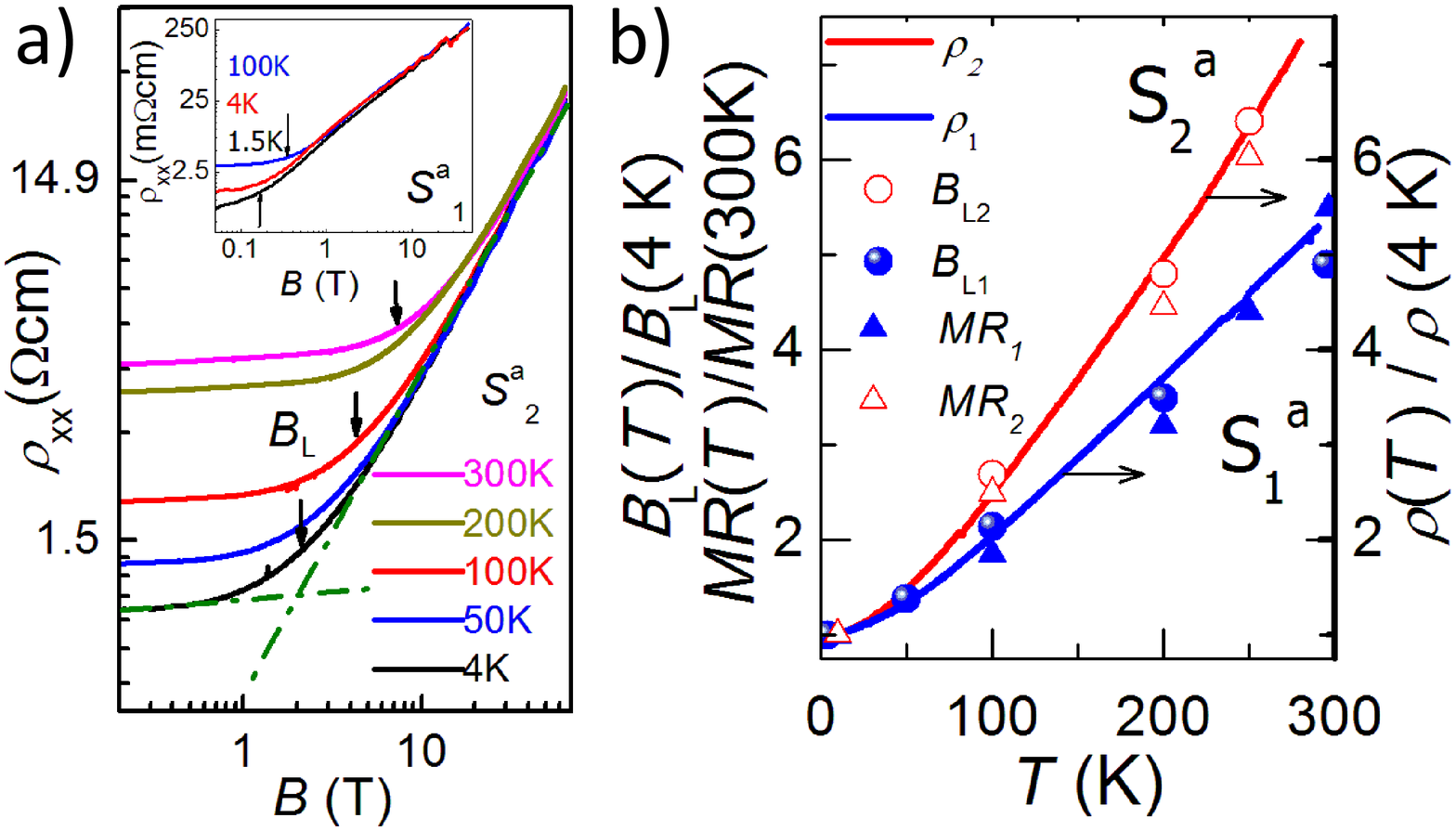}
    \includegraphics[width=8cm]{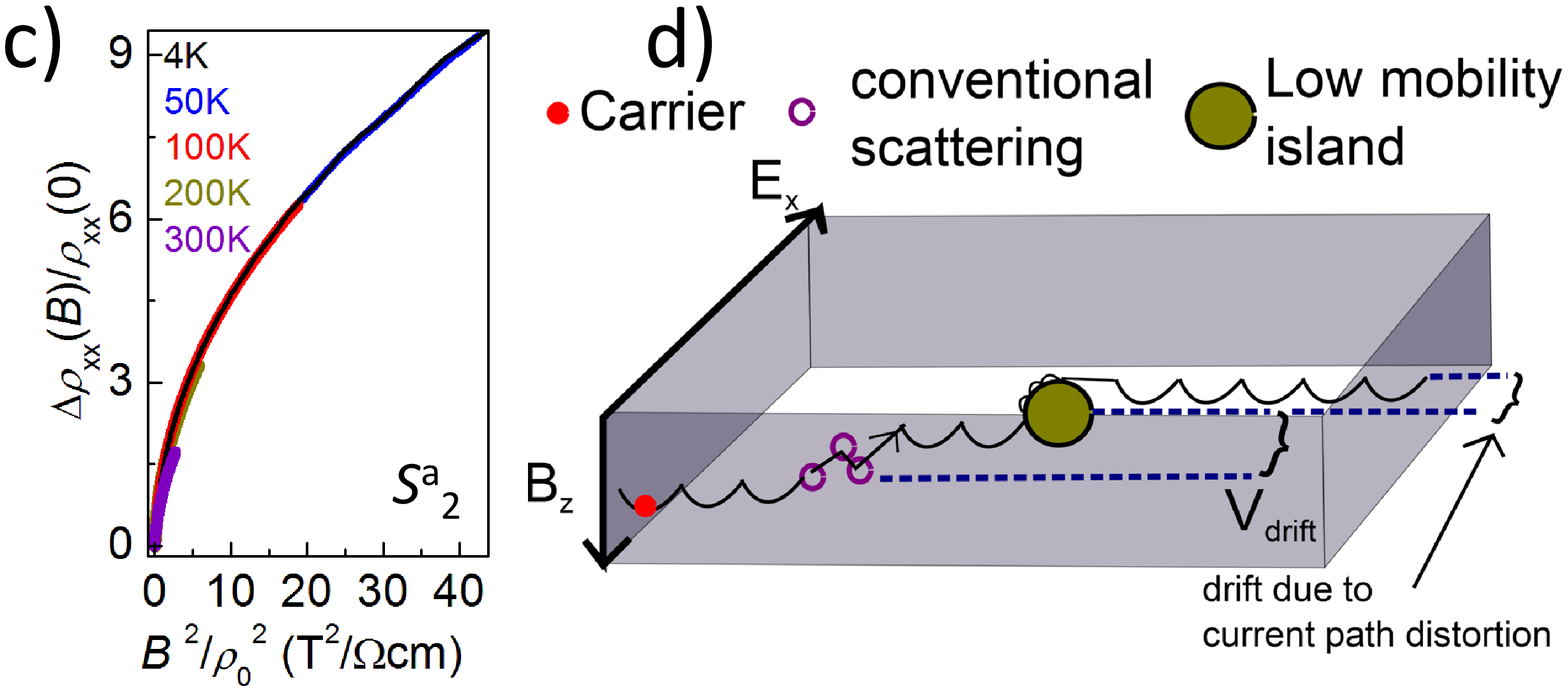}
  \caption{Linear MR and mobilities. a) Log-log plot of resistance versus field for $S^{a}_{2}$ and $S^{a}_{1}$ (inset). The crossover field, $B_L$, to the linear MR is indicated by arrows. b) The temperature dependence
of ratios of mobility, $\rho \sim \mu^-1_c$ (solid lines), $B_L$ (squares) normalized to the 4~K values, and the change in MR (triangles) show the same temperature dependence. c) Kohler's plots for $S^{a}_{2}$ showing the collapse of all magnetoresistance curves into one [from Fig. \ref{fig1}] below the Debye temperature, 200~K \cite{debye}. d) Schematic diagram of scattering processes in Cd$_{3}$As$_{2}$. }
  \label{fig3}
  \end{figure}

{\it Linear magnetoresistance}
Now we discuss the origin of the unconventional linear MR in a transverse magnetic field for two crystals
 of Cd$_3$As$_2$ (shown initially in Fig.\ref{fig1}a)) plotted in Fig.\ref{fig3}a)
 on a log-log scale to emphasize the low field behaviour. We observe
 that the linear MR behaviour
 is established above a crossover field, $B_{\rm L}$.
 Interestingly, we find that $B_{\rm L}$
 and the relative change in magnetoresistance, $MR=\Delta \rho_{xx}$(B)/$\rho_{xx}$(0),
  vary with temperature in the same ratio as the mobility, $\mu_c$
and, consequently, the resistivity ratio ($\rho \sim \mu^{-1}_c$)[see Fig.\ref{fig3}b)].
  Furthermore, we find that all MR curves
  collapse onto a single curve in a Kohler's plot for temperatures below 200~K, suggesting
   that a single relevant scattering process is dominant in Cd$_3$As$_2$,
as shown in Fig.\ref{fig3}c). Small deviations at higher temperatures are caused by the onset of phonon scattering,
 consistent with the Debye temperature of 200~K \cite{debye}.

%\begin{center}
\begin{table*}[htb]
\caption{Band parameters extracted from quantum oscillations, such as
 frequencies for two different orientations ($F_1$ for $B \parallel $ [112] axis and $F_2$
 for $B \perp $ [112]),
 %  the corresponding Dirac energy assuming linear dispersion $E_D$,
 Fermi velocities, $v_{\rm F}=\hbar k_F/m_{eff}$, the Berry's phase, $\beta$, the $g$-factor, the Dingle temperature, $T_{\rm D}$, the mean free path, $\ell$  and the quantum mobility, $\mu_{q}$.
  The carrier concentration, $n_{\rm SdH}$, was estimated assuming that the Fermi surface is a three-dimensional ellipsoid.
 The Hall effect data give the carrier concentration $n_{\rm Hall}$  and classical mobilities, $\mu_{c}$, and the mobility ratio,  $\mu_{c}/\mu_{q}$. The data are reported  for samples from different batches ($a$, $b$ and $c$) and they are compared to published data.}
\begin{tabular}{lccccccccccccc}
\hline
\hline
 & $F_1$  & $F_2$  &
 $n_{\rm SdH}$ & $n_{\rm Hall}$ & $m_{eff}$
 & $v_{\rm F}$
 & $T_{\rm D}$ & $\ell$ &
 $\mu_{q}$ & $\mu_{c}$ &
 $\mu_{c}/ \mu_{q}$ &   $g$& $\beta$ \\
 & T& T &
  10$^{18}$ cm$^{-3}$ & 10$^{18}$ cm$^{-3}$
  & $m_{e}$ &  10$^6$ m/s &  K & nm&
 m$^2$/Vs & m$^2$/Vs & & & $\pi $\\
\hline % inserts single horizontal line
$S^{a}_{1}$& 31(4) & 45(4) &  1.0(2)&1.8(2)& 0.023(4)&1.54(4)&15.4(8)&122(8)&0.60(1)&8.0(5)&13.3(4)&16(4)&0.83(8)\\ % inserting body of the table
$S^{b}_1$& 42(4) & 52(4) &  1.5(2)&2.5(2)& 0.031(3)&1.33(4)&14.4(8)&112(8)&0.47(1)&3.4(3)&7.1(4)&15(3)&1.08(6)\\
$S^{c}_1$& 67(4) & 74(4) &  3.1(2)&3.8(2)& 0.043(4)&1.21(4)&9.8(8)&150(8)&0.51(1)&2.9(3)&5.7(4)&-&0.84(4)\\
Lit. & 20-90 & 20-90 &  0.1-8&2-20& 0.03-0.08&0.4-12&11-17&-&0.1-$10^{4}$&1-$10^{3}$&1-$10^{4}$&2-100&- \\
Refs.& \cite{oldqo,luli} & \cite{oldqo}  & \cite{luli,onglmr} &\cite{luli,onglmr} &\cite{oldqo,luli} &\cite{oldqo,arpes1} &\cite{dingleb}&-&\cite{luli,onglmr}&\cite{onglmr}&\cite{onglmr}&\cite{stm,oldgfactor1,oldgfactor2}&- \\
\hline
\hline
\end{tabular}
\label{table1}
\end{table*}
%\end{center}

%A non-saturating and large linear MR to 65~T is unconventional.
The conventional MR
shows a quadratic dependence at low fields and saturation for Fermi surfaces
with closed orbits in high fields, such that $\mu_c B_L>1$; in our samples
the crossover field can be estimated as $B_L>$1~T.
Linear MR has been predicted by Abrikosov \cite{abrikosov}
to occur in the quantum limit,
only beyond the $n=1$ Landau level. However,
in our crystals the value of $B_L$ is  much lower
 than the position of the $n=1$ level around 32~T.

 Another explanation for the presence of linear MR has its origin in classical disorder models. For example,
 linear MR was realized for highly disordered
 % an extreme case of a two-dimensional random resistor network
\cite{parish,parish2}, or weakly disordered-high mobility samples  \cite{herring2}, thin films and quantum Hall systems \cite{simon}. The linear MR
arises because the local current density
acquires spatial fluctuations in both
magnitude and direction, as a result of the heterogeneity or
microstructure caused by
non-homogeneous carrier and mobility distribution [see Fig.\ref{fig3}d].
There are a series of experimental realization of linear MR in
disordered systems, such as Ag$_{2\pm \delta}$Se and Ag$_{2\pm \delta}$Te
\cite{Hu2008}, two-dimensional systems
(epitaxial graphite) \cite{Aamir2012,epitaxial},
In(As/Sb)\cite{inas}, %and Bi-based topological insulators \cite{biti},
LaSb$_2$ \cite{Budko1998}, LaAgSb$_2$ \cite{Wang2012}.

Monte Carlo simulations for a system with a few islands of enhanced scattering embedded in a medium of high mobility \cite{inas}, suggest that MR is linked to the generation of an effective drift velocity perpendicular to cycloid motion in applied electric field caused by multiple small angle scattering of charge carriers by the islands (see Fig.~\ref{fig3}d).
       For such a mechanism
      the mobility $\mu_c$ is determined by the island separation and
         depending on the value of $\frac{\delta \mu_c}{\mu_c}$, the linear MR emerging from this process will be associated with  $B_L \sim \mu_{c}^{-1}$, which tracks the island separation if $\frac{\delta \mu_c}{\mu_c} < 1$ and tracks $\delta \mu_c^{-1}$ if $\frac{\delta \mu_c}{\mu_c} > 1$.
         Thus, the absolute value of the linear MR and $B_L$ would
         vary like $\mu^{-1}_c$ (linked to $\rho$ values) [Fig.\ref{fig3}b]. This scaling
         is consistent with the classical disordered model
         originating from fluctuating mobilities for the observed linear MR in Cd$_3$As$_2$,

Lastly, we comment on the possible source of disorder in Cd$_{3}$As$_{2}$. STM measurements found disordered patches with a typical size of 10~nm and separated by distances of 50~nm, attributed to As vacancy clusters \cite{stm}, likely to appear during the growth in a Cd-rich environment with a small width formation for Cd$_3$As$_2$ \cite{cava}.
  Assuming a disorder density comparable to the carrier concentration, $n_{\rm SdH}$, and a dielectric constant of $\epsilon$=16 (see Ref.\cite{disordertheory}), one can estimate the classical mobility as being 30000~cm$^{2}$/Vs for Cd$_{3}$As$_{2}$,
 which is similar to our measured classical mobilities, $\mu_c$.
  The lower quantum mobility, $\mu_q$ corresponds to small angle scattering when carriers travel over the mean free path, $\ell  \sim 110 - 150$ nm, which is similar to the distribution of As vacancy clusters imaged by STM \cite{stm} [see Fig.\ref{fig3}d)]. Furthermore, a mobility ratio $\mu_c/\mu_q > 1$ points towards As vacancies as being the small angle scatterers in Cd$_{3}$As$_{2}$ \cite{dassharma}.
     Concerning the possible changes of the Fermi surface induced by magnetic field in in Cd$_3$As$_2$
  our data that access the quantum limit [for Sample $S^a_1$ in Fig.\ref{fig1}c)] we find
  no evidence of additional frequencies (only spin-splitting
  due to the large $g$ factors) or changes in scattering (Dingle term) up to 65~T.

In conclusion, we have used ultra high magnetic fields
to characterize the Fermi surface of Cd$_{3}$As$_{2}$ and
to understand the origin of its linear magnetoresistance.
The Fermi surface of Cd$_{3}$As$_{2}$
 has an elliptical shape with a non-trivial Berry's phase.
We find that the linear MR enhancement scales with mobility in Cd$_{3}$As$_{2}$
  and likely originates from fluctuating mobilities regions that caused inhomogeneous
  current paths.   Close to the quantum limit we find no evidence for Fermi surface reconstruction
 except the observed spin-splitting effects caused by the large $g$ factors.
  The robust sample dependent linear MR suggest a possible avenue for tuning sample quality and
 further enhancing its MR for useful practical devices.  %

We acknowledge fruitful discussions with John Chalker, Steve Simon, Zohar Ringel
and Gabor Halasz and useful comments given by David Macdougal and S. L. Bud'ko.
This work was mainly supported by EPSRC (EP/L001772/1, EP/I004475/1, EP/I017836/1).
Part of the work was performed at the LNCMI, member of the
European Magnetic Field Laboratory (EMFL).  AIC
acknowledges an EPSRC Career Acceleration Fellowship (EP/I004475/1).
Work done at Ames Lab was supported by the U.S. Department of Energy,
Office of Basic Energy Science, Division of Materials Sciences and Engineering. Ames Laboratory is
operated for the U.S. Department of Energy by Iowa State University under Contract No. DE-AC02-
07CH11358. Y.C. acknowledges the support from the EPSRC (UK) grant EP/K04074X/1 and
a DARPA (US) MESO project (no. N66001-11-1-4105).

\bibliography{cdas5}

\begin{thebibliography}{33}
\expandafter\ifx\csname natexlab\endcsname\relax\def\natexlab#1{#1}\fi
\expandafter\ifx\csname bibnamefont\endcsname\relax
  \def\bibnamefont#1{#1}\fi
\expandafter\ifx\csname bibfnamefont\endcsname\relax
  \def\bibfnamefont#1{#1}\fi
\expandafter\ifx\csname citenamefont\endcsname\relax
  \def\citenamefont#1{#1}\fi
\expandafter\ifx\csname url\endcsname\relax
  \def\url#1{\texttt{#1}}\fi
\expandafter\ifx\csname urlprefix\endcsname\relax\def\urlprefix{URL }\fi
\providecommand{\bibinfo}[2]{#2}
\providecommand{\eprint}[2][]{\url{#2}}

\bibitem[{\citenamefont{Wang et~al.}(2013)\citenamefont{Wang, Weng, Wu, Dai,
  and Fang}}]{theory}
\bibinfo{author}{\bibfnamefont{Z.}~\bibnamefont{Wang}},
  \bibinfo{author}{\bibfnamefont{H.}~\bibnamefont{Weng}},
  \bibinfo{author}{\bibfnamefont{Q.}~\bibnamefont{Wu}},
  \bibinfo{author}{\bibfnamefont{X.}~\bibnamefont{Dai}}, \bibnamefont{and}
  \bibinfo{author}{\bibfnamefont{Z.}~\bibnamefont{Fang}},
  \bibinfo{journal}{Phys. Rev. B} \textbf{\bibinfo{volume}{88}},
  \bibinfo{pages}{125427} (\bibinfo{year}{2013}).

\bibitem[{\citenamefont{Borisenko et~al.}(2014)\citenamefont{Borisenko, Gibson,
  Evtushinsky, Zabolotnyy, B\"uchner, and Cava}}]{arpes1}
\bibinfo{author}{\bibfnamefont{S.}~\bibnamefont{Borisenko}},
  \bibinfo{author}{\bibfnamefont{Q.}~\bibnamefont{Gibson}},
  \bibinfo{author}{\bibfnamefont{D.}~\bibnamefont{Evtushinsky}},
  \bibinfo{author}{\bibfnamefont{V.}~\bibnamefont{Zabolotnyy}},
  \bibinfo{author}{\bibfnamefont{B.}~\bibnamefont{B\"uchner}},
  \bibnamefont{and} \bibinfo{author}{\bibfnamefont{R.~J.} \bibnamefont{Cava}},
  \bibinfo{journal}{Phys. Rev. Lett.} \textbf{\bibinfo{volume}{113}},
  \bibinfo{pages}{027603} (\bibinfo{year}{2014}).

\bibitem[{\citenamefont{Madhab et~al.}(2013)\citenamefont{Madhab, SuYang,
  Sankar, Alidoust, Bian, Liu, Belopolski, Chang, Jeng, Lin et~al.}}]{arpes2}
\bibinfo{author}{\bibfnamefont{N.}~\bibnamefont{Madhab}},
  \bibinfo{author}{\bibfnamefont{X.}~\bibnamefont{SuYang}},
  \bibinfo{author}{\bibfnamefont{R.}~\bibnamefont{Sankar}},
  \bibinfo{author}{\bibfnamefont{N.}~\bibnamefont{Alidoust}},
  \bibinfo{author}{\bibfnamefont{G.}~\bibnamefont{Bian}},
  \bibinfo{author}{\bibfnamefont{C.}~\bibnamefont{Liu}},
  \bibinfo{author}{\bibfnamefont{I.}~\bibnamefont{Belopolski}},
  \bibinfo{author}{\bibfnamefont{T.-R.} \bibnamefont{Chang}},
  \bibinfo{author}{\bibfnamefont{H.-T.} \bibnamefont{Jeng}},
  \bibinfo{author}{\bibfnamefont{H.}~\bibnamefont{Lin}}, \bibnamefont{et~al.},
  \bibinfo{journal}{arXiv:1309.7892}  (\bibinfo{year}{2013}).

\bibitem[{\citenamefont{Liu et~al.}(2014)\citenamefont{Liu, Jiang, Zhou, Wang,
  Zhang, Weng, Prabhakaran, Mo, Peng, Dudin et~al.}}]{arpes3}
\bibinfo{author}{\bibfnamefont{Z.~K.} \bibnamefont{Liu}},
  \bibinfo{author}{\bibfnamefont{J.}~\bibnamefont{Jiang}},
  \bibinfo{author}{\bibfnamefont{B.}~\bibnamefont{Zhou}},
  \bibinfo{author}{\bibfnamefont{Z.~J.} \bibnamefont{Wang}},
  \bibinfo{author}{\bibfnamefont{Y.}~\bibnamefont{Zhang}},
  \bibinfo{author}{\bibfnamefont{H.~M.} \bibnamefont{Weng}},
  \bibinfo{author}{\bibfnamefont{D.}~\bibnamefont{Prabhakaran}},
  \bibinfo{author}{\bibfnamefont{S.-K.} \bibnamefont{Mo}},
  \bibinfo{author}{\bibfnamefont{H.}~\bibnamefont{Peng}},
  \bibinfo{author}{\bibfnamefont{P.}~\bibnamefont{Dudin}},
  \bibnamefont{et~al.}, \bibinfo{journal}{Nat. Mater.}
  \textbf{\bibinfo{volume}{13}}, \bibinfo{pages}{677 } (\bibinfo{year}{2014}).

\bibitem[{\citenamefont{Sangjun et~al.}(2014)\citenamefont{Sangjun, Brian,
  Andras, Benjamin, Itamar, Andrew, Quinn, Robert, Ashvin, and Ali}}]{stm}
\bibinfo{author}{\bibfnamefont{J.}~\bibnamefont{Sangjun}},
  \bibinfo{author}{\bibfnamefont{Z.}~\bibnamefont{Brian}},
  \bibinfo{author}{\bibfnamefont{G.}~\bibnamefont{Andras}},
  \bibinfo{author}{\bibfnamefont{F.}~\bibnamefont{Benjamin}},
  \bibinfo{author}{\bibfnamefont{K.}~\bibnamefont{Itamar}},
  \bibinfo{author}{\bibfnamefont{P.}~\bibnamefont{Andrew}},
  \bibinfo{author}{\bibfnamefont{G.}~\bibnamefont{Quinn}},
  \bibinfo{author}{\bibfnamefont{C.}~\bibnamefont{Robert}},
  \bibinfo{author}{\bibfnamefont{V.}~\bibnamefont{Ashvin}}, \bibnamefont{and}
  \bibinfo{author}{\bibfnamefont{Y.}~\bibnamefont{Ali}},
  \bibinfo{journal}{Nature Materials}  (\bibinfo{year}{2014}).

\bibitem[{\citenamefont{Yi et~al.}(2014)\citenamefont{Yi, Wang, Chen, Shi,
  Feng, Liang, Xie, He, He, Peng et~al.}}]{arpes4}
\bibinfo{author}{\bibfnamefont{H.}~\bibnamefont{Yi}},
  \bibinfo{author}{\bibfnamefont{Z.}~\bibnamefont{Wang}},
  \bibinfo{author}{\bibfnamefont{C.}~\bibnamefont{Chen}},
  \bibinfo{author}{\bibfnamefont{Y.}~\bibnamefont{Shi}},
  \bibinfo{author}{\bibfnamefont{Y.}~\bibnamefont{Feng}},
  \bibinfo{author}{\bibfnamefont{A.}~\bibnamefont{Liang}},
  \bibinfo{author}{\bibfnamefont{Z.}~\bibnamefont{Xie}},
  \bibinfo{author}{\bibfnamefont{S.}~\bibnamefont{He}},
  \bibinfo{author}{\bibfnamefont{J.}~\bibnamefont{He}},
  \bibinfo{author}{\bibfnamefont{Y.}~\bibnamefont{Peng}}, \bibnamefont{et~al.},
  \bibinfo{journal}{Sci. Rep.} \textbf{\bibinfo{volume}{4}}
  (\bibinfo{year}{2014}).

\bibitem[{\citenamefont{Feng et~al.}(2014)\citenamefont{Feng, Pang, Wu, Wang,
  Weng, Li, Dai, Fang, Shi, and Lu}}]{luli}
\bibinfo{author}{\bibfnamefont{J.}~\bibnamefont{Feng}},
  \bibinfo{author}{\bibfnamefont{Y.}~\bibnamefont{Pang}},
  \bibinfo{author}{\bibfnamefont{D.}~\bibnamefont{Wu}},
  \bibinfo{author}{\bibfnamefont{Z.}~\bibnamefont{Wang}},
  \bibinfo{author}{\bibfnamefont{H.}~\bibnamefont{Weng}},
  \bibinfo{author}{\bibfnamefont{J.}~\bibnamefont{Li}},
  \bibinfo{author}{\bibfnamefont{X.}~\bibnamefont{Dai}},
  \bibinfo{author}{\bibfnamefont{Z.}~\bibnamefont{Fang}},
  \bibinfo{author}{\bibfnamefont{Y.}~\bibnamefont{Shi}}, \bibnamefont{and}
  \bibinfo{author}{\bibfnamefont{L.}~\bibnamefont{Lu}},
  \bibinfo{journal}{arXiv:1405.6611}  (\bibinfo{year}{2014}).

\bibitem[{\citenamefont{Liang et~al.}(2014)\citenamefont{Liang, Quinn,
  Ali~Mazhar, Liu, Cava, and Ong}}]{onglmr}
\bibinfo{author}{\bibfnamefont{T.}~\bibnamefont{Liang}},
  \bibinfo{author}{\bibfnamefont{G.}~\bibnamefont{Quinn}},
  \bibinfo{author}{\bibfnamefont{N.}~\bibnamefont{Ali~Mazhar}},
  \bibinfo{author}{\bibfnamefont{M.}~\bibnamefont{Liu}},
  \bibinfo{author}{\bibfnamefont{R.~J.} \bibnamefont{Cava}}, \bibnamefont{and}
  \bibinfo{author}{\bibfnamefont{N.~P.} \bibnamefont{Ong}},
  \bibinfo{journal}{arXiv:1404.7794}  (\bibinfo{year}{2014}).

\bibitem[{\citenamefont{Ali et~al.}(2014)\citenamefont{Ali, Gibson, Jeon, Zhou,
  Yazdani, and Cava}}]{cava}
\bibinfo{author}{\bibfnamefont{M.~N.} \bibnamefont{Ali}},
  \bibinfo{author}{\bibfnamefont{Q.}~\bibnamefont{Gibson}},
  \bibinfo{author}{\bibfnamefont{S.}~\bibnamefont{Jeon}},
  \bibinfo{author}{\bibfnamefont{B.~B.} \bibnamefont{Zhou}},
  \bibinfo{author}{\bibfnamefont{A.}~\bibnamefont{Yazdani}}, \bibnamefont{and}
  \bibinfo{author}{\bibfnamefont{R.~J.} \bibnamefont{Cava}},
  \bibinfo{journal}{Inorganic Chemistry} \textbf{\bibinfo{volume}{53}},
  \bibinfo{pages}{4062} (\bibinfo{year}{2014}).

\bibitem[{\citenamefont{Canfield and Fisk}(1992)}]{Canfield1992}
\bibinfo{author}{\bibfnamefont{P.~C.} \bibnamefont{Canfield}} \bibnamefont{and}
  \bibinfo{author}{\bibfnamefont{Z.}~\bibnamefont{Fisk}},
  \bibinfo{journal}{Phil. Mag. B} \textbf{\bibinfo{volume}{65}},
  \bibinfo{pages}{1117} (\bibinfo{year}{1992}).

\bibitem[{\citenamefont{Blaha et~al.}(2001)\citenamefont{Blaha, Schwarz,
  Madsen, Kvasnicka, and Luitz}}]{Blaha2001}
\bibinfo{author}{\bibfnamefont{P.}~\bibnamefont{Blaha}},
  \bibinfo{author}{\bibfnamefont{K.}~\bibnamefont{Schwarz}},
  \bibinfo{author}{\bibfnamefont{G.}~\bibnamefont{Madsen}},
  \bibinfo{author}{\bibfnamefont{D.}~\bibnamefont{Kvasnicka}},
  \bibnamefont{and} \bibinfo{author}{\bibfnamefont{J.}~\bibnamefont{Luitz}},
  \emph{\bibinfo{title}{{WIEN2k, An Augmented Plane Wave + Local Orbitals
  Program for Calculating Crystal Properties}}} (\bibinfo{year}{2001}).

\bibitem[{\citenamefont{Shoenberg}(1984)}]{Shoenberg1984}
\bibinfo{author}{\bibfnamefont{D.}~\bibnamefont{Shoenberg}},
  \emph{\bibinfo{title}{Magnetic Oscillations in Metals}}
  (\bibinfo{publisher}{Cambridge University Press},
  \bibinfo{address}{Cambridge, England}, \bibinfo{year}{1984}).

\bibitem[{\citenamefont{Sharapov et~al.}(2004)\citenamefont{Sharapov, Gusynin,
  and Beck}}]{Sharapov2004}
\bibinfo{author}{\bibfnamefont{S.~G.} \bibnamefont{Sharapov}},
  \bibinfo{author}{\bibfnamefont{V.~P.} \bibnamefont{Gusynin}},
  \bibnamefont{and} \bibinfo{author}{\bibfnamefont{H.}~\bibnamefont{Beck}},
  \bibinfo{journal}{Phys. Rev. B} \textbf{\bibinfo{volume}{69}},
  \bibinfo{pages}{075104} (\bibinfo{year}{2004}).

\bibitem[{\citenamefont{Singh and Wallace}(1983)}]{oldgfactor1}
\bibinfo{author}{\bibfnamefont{M.}~\bibnamefont{Singh}} \bibnamefont{and}
  \bibinfo{author}{\bibfnamefont{P.}~\bibnamefont{Wallace}},
  \bibinfo{journal}{Solid State Communications} \textbf{\bibinfo{volume}{45}},
  \bibinfo{pages}{9} (\bibinfo{year}{1983}), ISSN \bibinfo{issn}{0038-1098}.

\bibitem[{\citenamefont{Wallace}(1979)}]{oldgfactor2}
\bibinfo{author}{\bibfnamefont{P.~R.} \bibnamefont{Wallace}},
  \bibinfo{journal}{physica status solidi (b)} \textbf{\bibinfo{volume}{92}},
  \bibinfo{pages}{49} (\bibinfo{year}{1979}), ISSN \bibinfo{issn}{1521-3951}.

\bibitem[{\citenamefont{Taskin and Ando}(2011)}]{Taskin2011}
\bibinfo{author}{\bibfnamefont{A.~A.} \bibnamefont{Taskin}} \bibnamefont{and}
  \bibinfo{author}{\bibfnamefont{Y.}~\bibnamefont{Ando}},
  \bibinfo{journal}{Phys. Rev. B} \textbf{\bibinfo{volume}{84}},
  \bibinfo{pages}{035301} (\bibinfo{year}{2011}).

\bibitem[{\citenamefont{Wright and McKenzie}(2013)}]{mckenzie}
\bibinfo{author}{\bibfnamefont{A.~R.} \bibnamefont{Wright}} \bibnamefont{and}
  \bibinfo{author}{\bibfnamefont{R.~H.} \bibnamefont{McKenzie}},
  \bibinfo{journal}{Phys. Rev. B} \textbf{\bibinfo{volume}{87}},
  \bibinfo{pages}{085411} (\bibinfo{year}{2013}).

\bibitem[{\citenamefont{Bartkowski et~al.}(1989)\citenamefont{Bartkowski,
  Pompe, and Hegenbarth}}]{debye}
\bibinfo{author}{\bibfnamefont{K.}~\bibnamefont{Bartkowski}},
  \bibinfo{author}{\bibfnamefont{G.}~\bibnamefont{Pompe}}, \bibnamefont{and}
  \bibinfo{author}{\bibfnamefont{E.}~\bibnamefont{Hegenbarth}},
  \bibinfo{journal}{physica status solidi (a)} \textbf{\bibinfo{volume}{111}},
  \bibinfo{pages}{K165} (\bibinfo{year}{1989}), ISSN \bibinfo{issn}{1521-396X}.

\bibitem[{\citenamefont{Rosenman}(1969)}]{oldqo}
\bibinfo{author}{\bibfnamefont{I.}~\bibnamefont{Rosenman}},
  \bibinfo{journal}{J. Phys. Chem. Sol.} \textbf{\bibinfo{volume}{30}},
  \bibinfo{pages}{1385} (\bibinfo{year}{1969}), ISSN \bibinfo{issn}{0022-3697}.

\bibitem[{\citenamefont{Blom et~al.}(1980)\citenamefont{Blom, Cremers, Neve,
  and Gelten}}]{dingleb}
\bibinfo{author}{\bibfnamefont{F.}~\bibnamefont{Blom}},
  \bibinfo{author}{\bibfnamefont{J.}~\bibnamefont{Cremers}},
  \bibinfo{author}{\bibfnamefont{J.}~\bibnamefont{Neve}}, \bibnamefont{and}
  \bibinfo{author}{\bibfnamefont{M.}~\bibnamefont{Gelten}},
  \bibinfo{journal}{Solid State Communications} \textbf{\bibinfo{volume}{33}},
  \bibinfo{pages}{69 } (\bibinfo{year}{1980}), ISSN \bibinfo{issn}{0038-1098}.

\bibitem[{\citenamefont{Abrikosov}(1998)}]{abrikosov}
\bibinfo{author}{\bibfnamefont{A.~A.} \bibnamefont{Abrikosov}},
  \bibinfo{journal}{Phys. Rev. B} \textbf{\bibinfo{volume}{58}},
  \bibinfo{pages}{2788} (\bibinfo{year}{1998}).

\bibitem[{\citenamefont{Parish and Littlewood}(2003)}]{parish}
\bibinfo{author}{\bibfnamefont{M.}~\bibnamefont{Parish}} \bibnamefont{and}
  \bibinfo{author}{\bibfnamefont{P.}~\bibnamefont{Littlewood}},
  \bibinfo{journal}{Nature} \textbf{\bibinfo{volume}{426}}
  (\bibinfo{year}{2003}).

\bibitem[{\citenamefont{Hu et~al.}(2007)\citenamefont{Hu, Parish, and
  Rosenbaum}}]{parish2}
\bibinfo{author}{\bibfnamefont{J.}~\bibnamefont{Hu}},
  \bibinfo{author}{\bibfnamefont{M.~M.} \bibnamefont{Parish}},
  \bibnamefont{and} \bibinfo{author}{\bibfnamefont{T.~F.}
  \bibnamefont{Rosenbaum}}, \bibinfo{journal}{Phys. Rev. B}
  \textbf{\bibinfo{volume}{75}}, \bibinfo{pages}{214203}
  (\bibinfo{year}{2007}).

\bibitem[{\citenamefont{Herring}(1960)}]{herring2}
\bibinfo{author}{\bibfnamefont{C.}~\bibnamefont{Herring}}, \bibinfo{journal}{J.
  Appl. Physics} \textbf{\bibinfo{volume}{31}}, \bibinfo{pages}{1939}
  (\bibinfo{year}{1960}).

\bibitem[{\citenamefont{Simon and Halperin}(1994)}]{simon}
\bibinfo{author}{\bibfnamefont{S.~H.} \bibnamefont{Simon}} \bibnamefont{and}
  \bibinfo{author}{\bibfnamefont{B.~I.} \bibnamefont{Halperin}},
  \bibinfo{journal}{Phys. Rev. Lett.} \textbf{\bibinfo{volume}{73}},
  \bibinfo{pages}{3278} (\bibinfo{year}{1994}).

\bibitem[{\citenamefont{Hu and Rosenbaum}(2008)}]{Hu2008}
\bibinfo{author}{\bibfnamefont{J.}~\bibnamefont{Hu}} \bibnamefont{and}
  \bibinfo{author}{\bibfnamefont{T.~F.} \bibnamefont{Rosenbaum}},
  \bibinfo{journal}{Nature Materials} \textbf{\bibinfo{volume}{7}},
  \bibinfo{pages}{697} (\bibinfo{year}{2008}).

\bibitem[{\citenamefont{Aamir et~al.}(2012)\citenamefont{Aamir, Goswami,
  Baenninger, Tripathi, Pepper, Farrer, Ritchie, and Ghosh}}]{Aamir2012}
\bibinfo{author}{\bibfnamefont{M.~A.} \bibnamefont{Aamir}},
  \bibinfo{author}{\bibfnamefont{S.}~\bibnamefont{Goswami}},
  \bibinfo{author}{\bibfnamefont{M.}~\bibnamefont{Baenninger}},
  \bibinfo{author}{\bibfnamefont{V.}~\bibnamefont{Tripathi}},
  \bibinfo{author}{\bibfnamefont{M.}~\bibnamefont{Pepper}},
  \bibinfo{author}{\bibfnamefont{I.}~\bibnamefont{Farrer}},
  \bibinfo{author}{\bibfnamefont{D.~A.} \bibnamefont{Ritchie}},
  \bibnamefont{and} \bibinfo{author}{\bibfnamefont{A.}~\bibnamefont{Ghosh}},
  \bibinfo{journal}{Phys. Rev. B} \textbf{\bibinfo{volume}{86}},
  \bibinfo{pages}{081203} (\bibinfo{year}{2012}).

\bibitem[{\citenamefont{Friedman et~al.}(2010)\citenamefont{Friedman, Tedesco,
  Campbell, Culbertson, Aifer, Perkins, Myers-Ward, Hite, Eddy, Jernigan
  et~al.}}]{epitaxial}
\bibinfo{author}{\bibfnamefont{A.~L.} \bibnamefont{Friedman}},
  \bibinfo{author}{\bibfnamefont{J.~L.} \bibnamefont{Tedesco}},
  \bibinfo{author}{\bibfnamefont{P.~M.} \bibnamefont{Campbell}},
  \bibinfo{author}{\bibfnamefont{J.~C.} \bibnamefont{Culbertson}},
  \bibinfo{author}{\bibfnamefont{E.}~\bibnamefont{Aifer}},
  \bibinfo{author}{\bibfnamefont{F.~K.} \bibnamefont{Perkins}},
  \bibinfo{author}{\bibfnamefont{R.~L.} \bibnamefont{Myers-Ward}},
  \bibinfo{author}{\bibfnamefont{J.~K.} \bibnamefont{Hite}},
  \bibinfo{author}{\bibfnamefont{C.~R.} \bibnamefont{Eddy}},
  \bibinfo{author}{\bibfnamefont{G.~G.} \bibnamefont{Jernigan}},
  \bibnamefont{et~al.}, \bibinfo{journal}{Nano Letters}
  \textbf{\bibinfo{volume}{10}}, \bibinfo{pages}{3962} (\bibinfo{year}{2010}).

\bibitem[{\citenamefont{Kozlova et~al.}(2012)\citenamefont{Kozlova, Mori,
  Makarovsky, Eaves, Zhuang, Krier, and Patan}}]{inas}
\bibinfo{author}{\bibfnamefont{N.}~\bibnamefont{Kozlova}},
  \bibinfo{author}{\bibfnamefont{N.}~\bibnamefont{Mori}},
  \bibinfo{author}{\bibfnamefont{O.}~\bibnamefont{Makarovsky}},
  \bibinfo{author}{\bibfnamefont{L.}~\bibnamefont{Eaves}},
  \bibinfo{author}{\bibfnamefont{Q.}~\bibnamefont{Zhuang}},
  \bibinfo{author}{\bibfnamefont{A.}~\bibnamefont{Krier}}, \bibnamefont{and}
  \bibinfo{author}{\bibfnamefont{A.}~\bibnamefont{Patan}},
  \bibinfo{journal}{Nature Communications} \textbf{\bibinfo{volume}{3}}
  (\bibinfo{year}{2012}).

\bibitem[{\citenamefont{Bud'ko et~al.}(1998)\citenamefont{Bud'ko, Canfield,
  Mielke, and Lacerda}}]{Budko1998}
\bibinfo{author}{\bibfnamefont{S.~L.} \bibnamefont{Bud'ko}},
  \bibinfo{author}{\bibfnamefont{P.~C.} \bibnamefont{Canfield}},
  \bibinfo{author}{\bibfnamefont{C.~H.} \bibnamefont{Mielke}},
  \bibnamefont{and} \bibinfo{author}{\bibfnamefont{A.~H.}
  \bibnamefont{Lacerda}}, \bibinfo{journal}{Phys. Rev. B}
  \textbf{\bibinfo{volume}{57}}, \bibinfo{pages}{13624} (\bibinfo{year}{1998}).

\bibitem[{\citenamefont{Wang and Petrovic}(2012)}]{Wang2012}
\bibinfo{author}{\bibfnamefont{K.}~\bibnamefont{Wang}} \bibnamefont{and}
  \bibinfo{author}{\bibfnamefont{C.}~\bibnamefont{Petrovic}},
  \bibinfo{journal}{Phys. Rev. B} \textbf{\bibinfo{volume}{86}},
  \bibinfo{pages}{155213} (\bibinfo{year}{2012}).

\bibitem[{\citenamefont{Brian}(2014)}]{disordertheory}
\bibinfo{author}{\bibfnamefont{S.}~\bibnamefont{Brian}},
  \bibinfo{journal}{arXiv:1406.2318}  (\bibinfo{year}{2014}).

\bibitem[{\citenamefont{Das~Sarma and Stern}(1985)}]{dassharma}
\bibinfo{author}{\bibfnamefont{S.}~\bibnamefont{Das~Sarma}} \bibnamefont{and}
  \bibinfo{author}{\bibfnamefont{F.}~\bibnamefont{Stern}},
  \bibinfo{journal}{Phys. Rev. B} \textbf{\bibinfo{volume}{32}},
  \bibinfo{pages}{8442} (\bibinfo{year}{1985}).

\end{thebibliography}

\end{document}